# The Link Between Packing Morphology and the Distribution of Contact Forces and Stresses in Packings of Highly Non-Convex Particles

N. A. Conzelmann[1,2], A. Penn[1], M. N. Partl[2], F. J. Clemens[2], L. D. Poulikakos[2], C. R. Müller[1]

[1] ETH Zürich, Laboratory of Energy Science and Engineering, Department of Mechanical and Process Engineering, Institute of Energy and Process Engineering, Leonhardstrasse 21, 8092 Zürich, Switzerland

[2] Empa – Swiss Federal Laboratories for Materials Science and Technology, Überlandstrasse 129, 8600 Dübendorf, Switzerland

Keywords: Packing, Particle Shape, Non-convex Particles, Force Distribution, Force Chains, Morphology

ABSTRACT – An external load on a particle packing is distributed internally through a heterogeneous network of particle contacts. This contact force distribution determines the stability of the particle packing and the resulting structure. Here, we investigate the homogeneity of the contact force distribution in packings of highly non-convex particles both in two-dimensional (2D) and three-dimensional (3D) packings. A newly developed discrete element method is used to model packings of non-convex particles of varying sphericity. Our results establish that in 3D packings the distribution of the contact forces in the normal direction becomes increasingly heterogeneous with decreasing particle sphericity. However, in 2D packings the contact force distribution is independent of particle sphericity, indicating that results obtained in 2D packings cannot be extrapolated readily to 3D packings. Radial distribution functions (RDFs) show that the crystallinity in 3D packings decreases with decreasing particle sphericity. We link the decreasing homogeneity of the contact force distributions to the decreasing crystallinity of 3D packings. These findings are complementary to the previously observed link between the heterogeneity of the contact force distribution and a decreasing packing crystallinity due to an increasing polydispersity of spherical particles.

## Introduction

Densely packed granular materials are frequently encountered in every-day life, for example, in civil engineering applications such as railway track ballast or porous asphalt pavements [1,2]. Such packings often undergo compaction either by forced compression or due to the material's own weight [3,4]. It has been well established that internal forces in packed granular materials are not distributed homogeneously [5,6], but instead, forces are transmitted within the material through a network of so-called force chains. This force chain network traverses through the whole particle packing, but transmits forces through only a subset of the packed particles that are subject to above-average loads [7,8]. When exceeding a critical stress value in a force chain, buckling and/or slipping events can occur which result in large scale rearrangements of the packing [9,10]. Hence, the topology of the force chain network affects critically the structural stability of the packing [11].

Studies on force chains can be largely categorized into studies that consider (i) 2D packings [6,7], i.e. packings comprised of only one particle layer, or 3D packings [5,12] and (ii) packings that are compressed or sheared [6,7,12]. Compressed packings are obtained either by the uniaxial compression of particles in a rigid-walled container via a moving piston [12,13], or by its isotropic compression via two perpendicular walls [6,14]. To shear stress packings, various methods are used. For example placing particles in a rectangular confinement and compressing in one direction while expanding in another direction [6]. Numerically, packings may be sheared by compressing the packing vertically, while introducing a constant horizonal velocity to a frictional top wall and allowing a free horizontal movement of the particles [14,15].

In 2D packings, force chains were first qualitatively and later quantitatively probed by transmitting polarized light through a sheared packing of photoelastic discs, visualizing the stress in the discs in the form of fringes [7,16]. This method was extended by Liu et al. [5] to 3D packings by immersing photoelastic beads in a liquid with a matching refractive index. It was found that the magnitude of the contact forces is distributed heterogeneously through the packing. To this date, there is no agreement on how to quantify force chains, but commonly the distribution of contact forces is quantified by the probability distributions of the contact forces [5,12,17]. For example, Liu et al. [5] placed carbon paper onto the inner surfaces of a container holding a particle packing. By calibrating the size of the imprints on the carbon paper against a known force they obtained the probability distribution ($P(f)$ with $f = F_n/<F_n>$) of the normal contact forces ($F_n$) normalized by the mean normal contact force ($<F_n>$). It was found that the probability of finding large normalized forces (i.e. $f > 1$) decays exponentially with increasing force magnitude. To explain this experimental observation, a theoretical model was proposed that assumes that the dominant mechanism which gives rise to force chains is governed by the heterogeneity of the granular packing, causing in turn, an unequal force distribution on the individual particles. It was observed further that $P(f)$ has a peak at $f = 1$ and that $P(f) \rightarrow 0$ for $f \rightarrow 0$. The shape of the observed probability distribution function of the contact forces resembles a characteristic shape commonly observed for disc- and sphere-shaped particles [5,12,13,15,17]. This characteristic shape is shown schematically in Figure 1 and labelled as type A, while the characteristic distribution labelled type B is, for example, observed in sheared packings of non-spherical particles [14,15]. A shortcoming of the carbon paper method is the difficulty to distinguish between beads that do not transmit a force and voids. This aspect was studied further by Mueth et al. [12] determining the fraction of contacts in compressed 3D packings that transmit forces that are sufficiently low to not leave an imprint on the carbon paper. Incorporating this additional information, Mueth et al. [12] found that $P(f)$ has a saddle point at $f = 1$ and $P(f)$ increases for $f \rightarrow 0$ instead of approaching zero as proposed by Liu et al. [5]. Hence, for $f > 0.5$ the $P(f)$ as observed by Mueth et al. [12] has a concave shape that is characteristic of a distribution of type A, but combined with an increasing probability for $f \rightarrow 0$, that is characteristic for a type B distribution (Figure 1). The previous observation of Liu et al. [5] that $P(f)$ decays exponentially for $f > 1$ was confirmed by Mueth et al. [12].

Currently, there is no agreement on the mathematical function that describes $P(f)$ best. While some authors have argued that in a packing of spheres (type A shown in Figure 1) $P(f)$ can be fitted best by a Gaussian distribution [13,18], Mueth et al. [12] proposed Eq. (1) since a saddle point rather than a peak (at $f \approx 1$) was observed.

$$P(f) = a(1 - be^{-f^2})e^{-\beta f} \qquad (1)$$

In this equation $a$, $b$ and $\beta$ are fitting parameters. However, as shown in Figure 1, neither a Gaussian function nor Eq. (1) describe force distributions that have a type B shape. Nonetheless, it is generally agreed that for $f > 1$, $P(f)$ decays exponentially, independent of the domain dimensionality and particle sphericity. This observation motivated Azéma and Radjai [15] to propose the following fit to the tail ($f > 1$) of $P(f)$:

$$P(f) = e^{-kf}, \qquad f > 1 \qquad (2)$$

Here, the exponent $k$ is a fitting parameter that will be used to quantify the length of the exponential tail of $P(f)$.

Majmudar and Behringer [6] further improved the quantification of the magnitude of contact forces by using photoelastic discs and acquiring high-resolution photographs of 2D packings, allowing them to distinguish the individual interference fringes in the discs. Solving an inverse problem, which relates the number of fringes observed in a disk to the magnitude of the contact forces, the normal and tangential contact forces at each contact point were determined. Using this improved experimental technique, it was observed that in sheared packings fewer particles transmit large forces compared to compressed packings leading to more distinct force chains. Furthermore, it was shown that $P(f)$ in sheared and compressed packings of discs resembles a type A distribution with a peak at $f \approx 1$. However the coefficient $k$ of the exponential decay is smaller for sheared packings when compared to compressed packings.

Despite the continuous development and improvement of experimental techniques to visualize and quantify contact forces, it remains challenging to extract quantitative information of contact forces, in particular in 3D packings that are of high practical relevance. To address these challenges, the discrete element method (DEM) has established itself as an alternative to experimental approaches [19], providing detailed information on force networks in granular systems [20]. For example, Luding [17] used the DEM to investigate how the spatial stress distribution changes if polydispersity is introduced into packings of discs organized in a perfectly hexagonal lattice. For exactly monodisperse particle packings, particle stresses are distributed uniformly, in agreement with the hypothesis of Liu et al. [5]. However, as soon as polydispersity is introduced by varying the diameter (as little as ±0.33% of the mean diameter) a heterogeneous stress distribution, i.e. the occurrence of force chains, was observed [17]. Both experimental and numerical works [5,17] have established a link between the packing morphology and the contact force distribution in packings of spherical particles. However, thus far, this link has not been investigated for non-spherical particles.

Among the non-spherical particle packings studied, those composed of highly non-convex particles are particularly interesting as non-convex particles can interlock, forming packings that may sustain compressive and tensile forces despite containing purely non-cohesive particles [13,21,22]. Owing to these particular characteristics, packings of highly non-convex, interlocking particles may find practical relevance, for example in architecture by enabling novel construction concepts such as aleatory construction [21,23–25]. However, despite their intriguing characteristics, so far, only a few studies have investigated the distribution of contact forces, $P(f)$, in packings composed of non-spherical particles. For example, Gan et al. [26] performed 3D DEM simulations of packings of oblate ellipsoids with their sphericity ($\Psi$) varying between 1 to 0.7. Interestingly, the $P(f)$ for ellipsoids was similar to the distribution of spheres, i.e. $P(f)$ peaks close to $f = 1$ and for $f > 1$ $P(f)$ decays exponentially (type A distribution). The exponential decay was fitted by *Eq. (2)* with $k$ ranging between 1.2 and 1.4 depending on particle sphericity. However, there did not seem to be a clear correlation between the sphericity of the particles and the exponent $k$ characterizing the decay. Similar results were reported by Saint-Cyr et al. [14] who simulated compressed packings of particle clusters composed of three discs glued together in a triangular arrangement (trimers). The sphericity of the trimer particles was varied between 1 and 0.76 by varying the overlap of the trimer particles. A key finding of their work was that in compressed packings of trimers, $P(f)$ resembles the distribution of spheres (type A) with an exponential decay (for $f > 1$) with $k = 1.7$, independent of the particle sphericity, hence confirming the results of Gan et al. [26]. When the compressed trimer particles were also sheared (instead of only compressed), $k$ decreased with decreasing sphericity ($\Psi$), i.e. $k$ reduces from 1.7 to 1 for $\Psi$ decreasing from 1 to 0.76. The behaviour of the reference case (discs with $\Psi$ = 1) was different in that $P(f)$ was not affected by the addition of shear. Furthermore, Saint-Cyr et al. [14] showed that in sheared packings of spheres, the shape of $P(f)$ is concave and resembles a type A distribution. However, for non-spherical particles with $\Psi < 0.96$, $P(f)$ increases for $f \to 0$, leading to a type B distribution without a peak. Further, it was found that $\lim_{f \to 0} P(f)$ increased with decreasing sphericity. The decreasing value of $k$ for decreasing particle sphericity and the absence of a peak at $f \approx 1$ leads to the key conclusion that in sheared packings $P(f)$ becomes increasingly heterogeneous for decreasing sphericity. The results of Saint-Cyr et al. were confirmed by Azéma and Radjai [15] in 2D simulations of sheared, half-disc-capped rectangular particles which resemble 2D spherocylinders. Azéma and Radjai [15] varied the sphericity of the particles from 1 to 0.82 and found the exponent $k$ in $P(f) = e^{-kf}$ to decreases from 1.8 to 0.85, respectively. Moreover, they could also confirm that for particles with $\Psi < 0.99$ $P(f)$ resembles a type B distribution where $\lim_{f \to 0} P(f)$ increased for decreasing $\Psi$. Highly non-convex particles of very low sphericity ($\Psi$ = 0.45) (and spheres as a reference case) were studied by Murphy et al. [13]. The objective of their work was to find particle shapes that can form free-standing, externally unconfined, packings that can support load (i.e. searching for packings of interlocking particles that can sustain compressive and tensile stresses). Particles of low sphericity were modeled by gluing together multiple spheres to yield Z-shaped particles. For the reference case, a 3D compressed packing of spheres, the well-established type A contact force distribution was observed, with a decay exponent of $k$ = 1.4. However, compressed 3D packings of Z-shaped particles have a force distribution of type B, similar to the distributions observed by Saint-Cyr et al. [14] and Azéma and Radjai [15] in 2D sheared packings of non-spherical particles (0.96 > $\Psi$ > 0.76). Additionally, the contact force distribution of Z-shaped particles ($\Psi$ = 0.45) had a very long exponential tail with an exponent $k$ ranging between 0.56 and 0.76 depending on the specific Z-shape.

From the above, one can conclude that in sheared 2D packings the exponential tail of the contact force distribution, $P(f)$, becomes longer with increasing particle non-sphericity, i.e. $k$ decreases with decreasing $\Psi$. In addition the shape of $P(f)$ transitions from type A shape to a type B upon shearing [14,15]. However, in compressed 2D packings a decrease of $k$, as well as a change from a type A to type B distribution, with increasing particle non-sphericity does not occur for particles with $\Psi > 0.76$. Conversely, in compressed 3D packings of particles with $\Psi$ = 0.45 a type B force distribution with k ≤ 0.76 was observed. Hence, it remains still unclear whether (i) contact force distributions of type A prevail in compressed 2D packings of particles of low sphericity ($\Psi < 0.76$) and (ii) the contact force distribution of low sphericity particles ($\Psi < 0.76$) changes from a type A to type B distribution when transitioning from 2D to 3D packings.

In this work we address these two questions to establish a general correlation between particle sphericity and the shape of the contact force distribution in compressed 2D and 3D packings of non-spherical particles. While 2D packings do not occur in nature they are frequently studied, in particular experimentally. To allow a comparison between the results of this work and previous experimental 2D and numerical 3D studies, packings of various dimensionality are studied here. In addition, we probe whether the conclusion drawn by Liu et al. [5] and Luding [17] for spherical particles, viz. that a more heterogeneous packing morphology leads to a longer exponential tail in the distribution function of the contact force, can be extended to non-spherical particle packings.

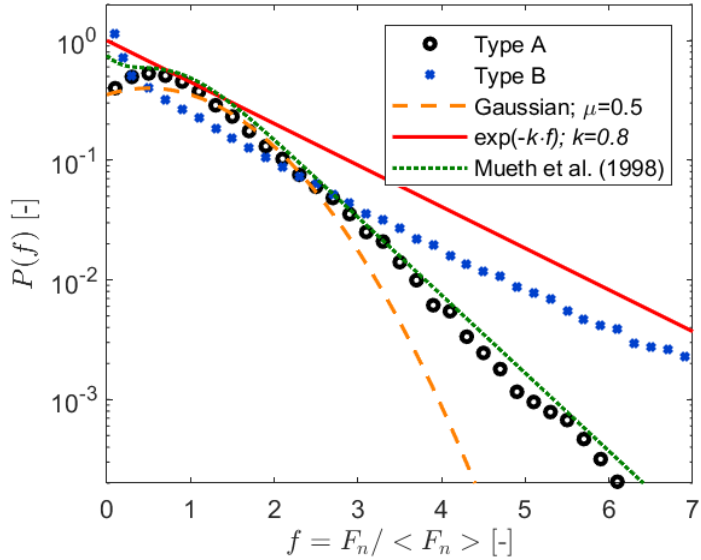

*Figure 1: Probability distributions of f, i.e. the normal contact force ($F_n$) normalized by the mean normal contact force (<$F_n$>). Distributions of type A and B represent characteristic shapes that have been observed in various particle packings. The type A and type B distributions were obtained for 3D packings studied in this work. Specifically, a type A distribution was obtained for spheres and a type B distribution for star-shaped particles with a sphericity (Ψ) equal to 0.419.*

## Methods

The DEM algorithm used in this study is a modification of the original concept proposed by Cundall and Strack which was developed for disc-shaped particles [20]. The present work considers particles that are created by combining multiple spherocylinders (cylinders with hemispherically capped ends), analogous to the commonly applied glued-sphere approach [27]. A spherocylinder is a computationally benevolent particle shape since all points on its surface have the same distance from the central axis (see Figure 2). The general concept used in the DEM to track particles and particle contacts has been well documented in the literature [20,27–32]. Hence, the following will only describe the contact model and the contact detection algorithm between spherocylinders.

### Particle contacts

Since all points on the surface of a spherocylinder have the same distance from the central axis (red dashed line in Figure 2), the contact detection for spherocylinders can be reduced to the task of finding the closest points between two line segments. We solve this task using the algorithm proposed by Lumelsky [33]. The point at which the contact forces act is called the contact point and corresponds to the center point (green point in Figure 2) of the line that connects the two closest points on each of the central axes (blue line in Figure 2). The normal contact force ($F_n$ in Figure 2) acts perpendicular to the surface of the particle, while the tangential force ($F_t$), resulting from friction, acts parallel to the surface. The direction in which the tangential force acts depends on the relative velocity of the particles at the contact point and the accumulated displacement between the particles.

If the angle between the central axes of the two contacting spherocylinders is less than two degrees, the contact is treated as a parallel contact (Figure 2b). The value of 2° is chosen as a feasible and efficient cut-off value based on additional numerical experiments. These experiments demonstrated that varying the cut-off angle between 0.01° and 5° does not affect the packing density nor the particle orientations. For a parallel contact, the middle of the parallel sections that align is chosen as the contact point (Figure 2b).

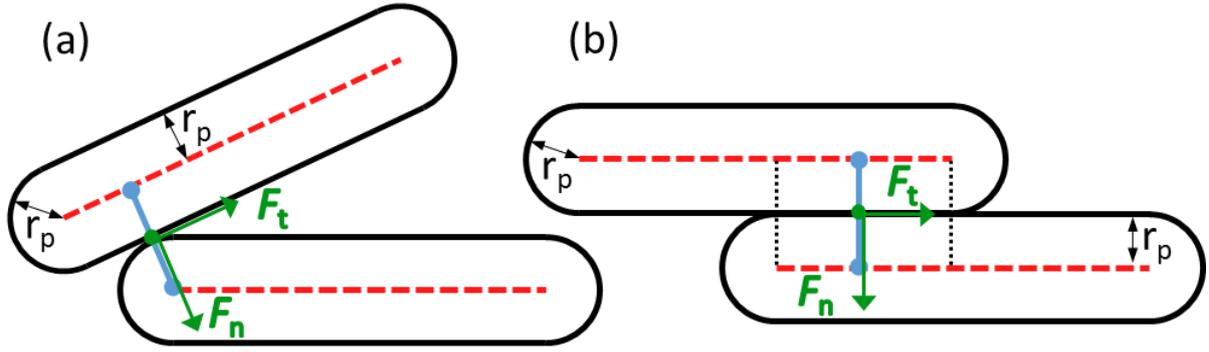

*Figure 2: Schematic of two contacting spherocylinders. The red dashed lines denote the central axis of each spherocylinder. The spherocylinder radius is labelled $r_p$. The blue line depicts the shortest distance between the central axis of the two spherocylinders. $F_n$ and $F_t$ depict the normal and tangential forces acting at the contact point (green): a) contact between a cylindrical section and a hemispherical end cap, b) parallel contact between spherocylinders.*

The contact between two particles is modeled by a linear spring-dashpot. The linear spring-dashpot model leads to a constant coefficient of restitution, independent of the collision velocity [34]. A linear model instead of a more realistic non-linear Hertzian model is justified as previous works have shown that packings of spherocylinders, when using either linear or Hertzian contact models, did not differ in the probability distribution of the contact forces [35]. It has also been argued that the material stiffness should increase with increasing contact area, e.g. in the case of a parallel contact between two spherocylinders (Figure 2b) [36–38]. However, a recent study has shown that varying the contact stiffness in the case of parallel contacts has little influence on the force distribution and structure in packings of spherocylinders [35]. Consequently, in this work, the normal and tangential stiffness are assumed to be constant, regardless of the geometry of the contact.

The contact force in the normal direction, $F_n$, between two contacting particles $i$ and $j$ is:

$$F_n = \max\left(0, \quad \frac{k_n}{2}\delta_n - \eta_n\sqrt{2m_{ij}k_n}v_n\right) \quad (3)$$

Here $k_n$ is the normal stiffness of the particles, $\delta_n$ is the overlap between the contacting particles, $\eta_n$ is the damping factor in the normal direction and $v_n$ is the normal component of the relative velocity between the particles at the point of contact. The effective inertial mass $m_{ij}$ is calculated by:

$$m_{ij} = \frac{m_i * m_j}{m_i + m_j} \quad (4)$$

In the tangential direction (subscript $t$), the maximal contact force is limited by Coulomb's law of friction and is calculated by:

$$F_t = \min\left(\mu\frac{k_n}{2}\delta_n, \quad \frac{k_t}{2}\delta_t - \eta_t\sqrt{2m_{ij}k_t}v_t\right) \quad (5)$$

where $\mu$ is the coefficient of friction, $\eta_t$ is the tangential damping factor, $k_t$ is the tangential stiffness and $v_t$ is the tangential component of the relative velocity between the particles at the point of contact. The accumulated tangential displacement at the contact is calculated as $\delta_t = \int v_t \, dt$.

### Simulation parameters

The density of the particles is fixed to $\rho$ = 1000 kg/m³ which is close to the density of many plastics, e.g. polyethylen (PE) or acrylonitrile butadiene styrene (ABS). Such plastics are commonly used in manufacturing techniques such as injection molding or fused deposition modelling, which can be employed to manufacture non-spherical particles. To choose a value for the normal particle stiffness $k_n$ two competing factors have to be considered: One factor is that a decreasing $k_n$ leads to a larger time step *dt* which decreases the computational time. As a rule of thumb *dt* should be at least 20 times smaller than the duration of a collision $t_{col}$ which can be approximated by a damped harmonic oscillator [39]:

$$t_{col} = \frac{\pi}{\sqrt{\frac{k_n}{m_{ij}}(1-\eta_n^2)}} \tag{6}$$

Furthermore, particles in a packing experience the weight of the particles above them which can lead to a vertical gradient of the magnitude of the contact forces. The influence of this gradient on the contact force probability distribution needs to be considered. Some researchers have eliminated the influence of this gradient by normalizing the contact forces by the weight of the particles above a given depth in the packing [5]. In our work a different approach is used to eliminate the effect of the vertical gradient in the magnitude of the contact forces that is by exerting an additional vertical load onto the packing through a forced compression at the top. If the contact forces due to the external compression are sufficiently high, the influence of the vertical gradient can be neglected. However, increasing the contact force between particles, without changing the stiffness, increases the overlap $\delta_n$ between the particles which can lead to computational stability problems. Hence, a sufficiently high $k_n$ is desired to be able to neglect the vertical gradient in the magnitude of the contact forces. For the given particle density of $\rho$ = 1000 kg/m³ an external compression force of 500 N was found, through simulations, to be sufficient to eliminate the influence of the vertical gradient in the magnitude of the contact forces. For an external compression force of 500 N a normal particle stiffness $k_n$ of 100'000 N/m is required to limit the particle overlaps to $\delta_n$ < 0.05 × $d_p$, where $d_p$ is the spherocylinder diameter. Additional simulation results show that increasing $k_n$ by an order of magnitude does not influence the contact force distributions; decreasing $k_n$ by an order of magnitude while keeping the compression force constant leads to unstable simulations. In conclusion, any set of values for the parameters $\rho$, $k_n$, *dt* and the compression force can be chosen, without affecting the results, as long as the set satisfies the conditions outlined above.

The damping factor in the normal direction $\eta_n$ is assumed to be constant and is related to the coefficient of restitution $\varepsilon$ via:

$$\varepsilon = exp\left(\frac{-\pi\eta_n}{\sqrt{1-\eta_n^2}}\right) \tag{7}$$

In this work, $\varepsilon$ = 0.53 ($\eta_n$ = 0.2) is chosen as previous studies have shown that varying $\varepsilon$ in the range 0.2 – 0.9 has little influence on the behaviour of dynamic granular systems such as rotating drums and flows down an inclined plane [40,41]. We expect the influence of $\varepsilon$ on the results of a static granular packing to be even less significant than for dynamic systems. For the tangential stiffness $k_t$ = 0.5 × $k_n$ is chosen in accordance with previous works [30,31,42], while for the friction coefficient a value of $\mu$ = 0.5 is chosen. A discussion about the influence of particle friction on the contact force probability distribution can be found in the Supplemental Material [43].

Friction at the domain walls is neglected by setting the coefficient of friction between particles and the domain walls to $\mu_w$ = 0, as some authors [44] have argued that friction between particles and the domain wall leads to a more heterogeneous particle packing. The influence of a frictional wall on a packing is most pronounced close to the wall. In the work reported here, the dimensionality of the domain is varied. Hence, to isolate the effect of dimensionality on a particle packing, the effect of wall friction has to be eliminated. Furthermore, other numerical works on non-spherical particles [13,14] have also chosen to neglect wall friction. For these reasons wall friction is also neglected in this work. Of course, frictionless walls are typically not observed in experiments. To aid comparison with experimental work the influence of the wall friction on the particle packing is discussed in the Supplemental Material [43].

Table 1 summarizes the values of the parameters used in the simulations. The value of the parameters of the confining walls are identical to those of the particles, except for the friction coefficient.

*Table 1: Material parameters used in the simulations*

| Parameter | Symbol | Value |
|---|---|---|
| Density | $\rho$ | 1000 kg/m³ |
| Normal stiffness | $k_n$ | 100'000 N/m |
| Tangential stiffness | $k_t$ | 50'000 N/m |
| Coefficient of restitution | $\varepsilon$ | 0.53 |
| Normal damping factor | $\eta_n$ | 0.2 |
| Tangential damping factor | $\eta_t$ | 0.2 |
| Coefficient of friction | $\mu$ | 0.5 |
| Wall friction coefficient | $\mu_w$ | 0 |
| Spherocylinder diameter | $d_p$ | 0.005 m |
| Time step | $dt$ | $2 \times 10^{-6}$ 1/s |

## Cluster particles

The combination of several spherocylinders to spherocylinder-cluster particles, analogous to the glued sphere approach [27,37], does not require additional contact detection routines [27]. The contact between two cluster particles can be treated as a contact between individual spherocylinders. The contact forces acting on the different spherocylinders belonging to a cluster are summed up and act on the center of gravity of the cluster.

In this study, two different types of cluster particles are investigated. In 2D packings, cross-shaped particles are used (Figure 3a-e). Cross-shaped particles are formed by intersecting two perpendicular spherocylinders of equal length in their centers. In simulations of pseudo-2D and 3D packings, star-shaped particles (also referred to as jacks or hexapods) are used. Such particles are formed by intersecting a cross-shaped particle with a third spherocylinder (of the same length) perpendicular to both spherocylinders that form the cross (Figure 3f). These particle shapes are chosen as they model non-convex geometries with a high order of symmetry and are easy to construct.

The non-convexity of cross- and star-shaped particles increases with decreasing sphericity ($\Psi$). Various definitions for sphericity have been proposed [45–47], whereby the most common definition is the ratio of the surface area of a sphere to the surface area of a non-spherical particle with the same volume as the sphere [46,48,49]. The present work uses this definition and thus for a sphere $\Psi = 1$ and for non-convex particles $\Psi < 1$. The sphericity of the particle shapes modeled in this work ranges from $\Psi = 0.99$ to 0.42, hence covering a broad range of shapes from almost sphere-like to very slender highly non-convex shapes (Figure 3e).

Some authors [50,51] have argued that several parameters, such as sphericity, blockiness and convexity, have to be specified to distinguish between different non-spherical particle shapes. We agree with the general rationale behind this proposal, however, this work investigates only two different shapes of particles, i.e. cross-shaped and star-shaped particles. For these two particle shapes any non-sphericity-describing parameter, e.g. blockiness or convexity, scales monotonically with (non-)sphericity. We believe that due to this monotonic scaling, the introduction of a second non-sphericity parameter would be redundant, provided that particles of one shape are compared only to particles of the same shape (as is the case of this study). An exception is made for the particle aspect ratio (defined as the ratio of the overall length of a particle $L$ to the diameter of the protruding arms $d_p$

(Figure 3)) which was used by other works [21,22] to describe star-shaped particles. To aid the comparison between our work and works [21,22] the relationship between the particle aspect ratio and the particle sphericity is given in the Supplemental Material [43].

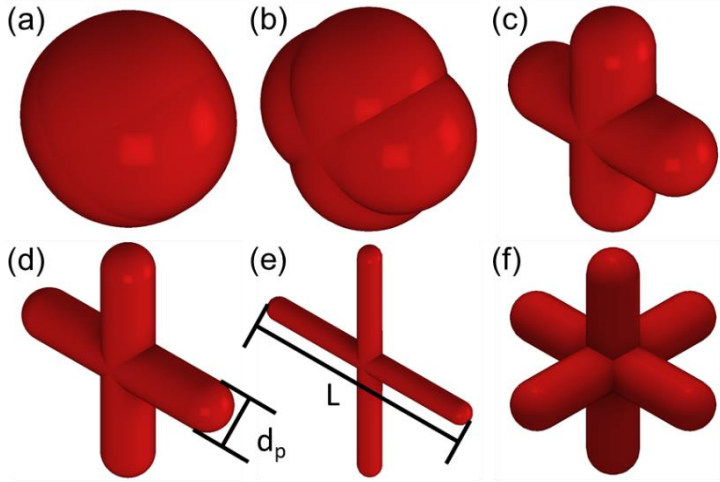

Figure 3: Selection of cross- and star-shaped particle geometries with varying sphericities (Ψ): a) Cross-shaped particle with Ψ=0.99, b) cross-shaped particle with Ψ=0.96, c) cross-shaped particle with Ψ=0.75, d) cross-shaped particle with Ψ=0.59, e) cross-shaped particle with Ψ=0.47 and f) star-shaped particle with Ψ=0.53.

## Simulation domains

Four different domain configurations (defined by the confining walls) are assessed in this work: (i) 2D simulations in which cross-shaped particles cannot move in the z-direction (coordinate system shown in Figure 4). In these 2D simulations the domain width $W_{dom}$ (x-direction) is equal to 30 times the particle length *L*. Therefore, $W_{dom}$ changes with particle shape Ψ; (ii) Pseudo-2D simulations in a cuboidal domain with a transverse thickness *T* (z-direction) equal to *L*. As in the 2D simulations, $W_{dom}$ = 30 × *L*; (iii) Pseudo-2D simulations in a cuboidal domain with *T* = 2 × *L* and $W_{dom}$ = 30 × *L* and (iv) 3D simulations in a cylindrical domain of diameter $D_d$ = 10 × *L*. A visualization of the pseudo-2D simulation with *T* = 2 × *L* is given in Figure 4a.

For 3D domains two diameters $D_d$ are investigated, i.e. $D_d$ = 6 × *L* and $D_d$ = 10 × *L*; for these two domain sizes very similar results are obtained for the contact force distributions. To have a larger data set for the contact force distributions, the results of the larger 3D domain ($D_d$ = 10 × *L*) are presented in the following. For the pseudo-2D domain, a transverse thickness of *T* = *L* is chosen; this thickness is as close as possible to a 2D domain while still allowing rotations around any given axis of the star-shaped particles. To investigate how a change in the transverse thickness affects the structure of the packing and transmission of contact forces, pseudo-2D domains with *T* = 2 × *L* are simulated. As the results of the pseudo-2D domains with *T* = 2 × *L* are similar to the 3D results, *T* is not increased further.

To initialize a simulation, the domain is filled by placing all of the particles on a regular lattice with a space of 1.5 × $d_p$ between each particle. The height of the simulation domain (y-direction) is chosen just large enough to accommodate all of the particles in the initialization lattice. The particles are initialized with a random rotational orientation and a random velocity *v* (-0.25 m/s < *v* < 0.25 m/s) in the y- and z-directions (2D and *T* = *L* cases) or a random velocity in all three directions (3D and *T* = 2 × *L* cases). The particles are allowed to settle for a time $t_{settle} = 3 * \sqrt{2h/g}$, where g is the acceleration due to gravity and *h* is the domain height. After time $t_{settle}$, the particles have come to rest as the average displacement of the particles per time-step approaches numeric precision. Once the particles have settled, the domain is compressed by moving a planar wall from the top downwards with a speed of 0.25 m/s until the packing exerts the external compression force of 500 N onto the top wall. This procedure simulates the uniaxial compression of a granular material in a container with rigid walls.

The number of particles in each simulation $N$ is such that the height of the packing (after compression) is at least twice as high as $W_{dom}$ (or in the case of a cylindrical domain the diameter of the cylinder). In the different configurations $N$ ranges from 1800 to 34500.

To avoid crystallization at low values of $\Psi$, a particle size distribution, is introduced. In 2D simulations, the polydispersity factor (by which the particle size is scaled) is ± 0.3 for $\Psi$ > 0.88, and ± 0.15 for 0.88 ≥ $\Psi$ > 0.75. For pseudo-2D and 3D simulations, the polydispersity factor is ± 0.2 for $\Psi$ > 0.86. As reference cases, packings of spherical particles (diameter $d_p$ = 0.005 m, polydispersity factor ± 0.2) in a 2D domain with $W_{dom}$ = 30 × $d_p$ and in a cylindrical domain with a cylinder diameter of 10 × $d_p$ are simulated.

### Data analysis: Contact forces, stress analysis and packing structure

Previous studies using spherical particles typically focused on the normal component of the contact force [12–14]. Thus, we likewise report here the distribution of the normal contact forces. For the computation of the probability distribution of the contact forces, $P(f)$, the normalized forces $f$ are sorted into 50 bins of size 0.2 (range 0-10).

Particle stresses are obtained by calculating first the stress tensor of the individual particles according to [52]:

$$\bar{\bar{\sigma}} = \frac{1}{V}\begin{bmatrix} \sigma_{xx} & \sigma_{xy} & \sigma_{xz} \\ \sigma_{yx} & \sigma_{yy} & \sigma_{yz} \\ \sigma_{zx} & \sigma_{zy} & \sigma_{zz} \end{bmatrix} = \frac{1}{V}\sum_c \bar{F}^c_{n+t} \bar{r}^c \qquad (8)$$

where $V$ is the particle volume, $c$ is the number of all particle contacts, $\bar{F}_{n+t}$ is the sum of the normal contact force and the tangential contact force and $\bar{r}$ is the vector pointing from the center of the particle to its contact point. The particle stress tensor is diagonalized to obtain the principal stresses. Additionally, this work reports the sum of the elements of the diagonalized stress matrix (trace) of each particle, i.e. the first stress invariant of the tensor for each particle ($I_{1,i}$). Similar to the presentation of the contact forces $I_{1,i}$ is normalized by its mean ($<I_1> = \frac{1}{N}\sum_i^N I_{1,i}$) yielding $i = I_{1,i}/<I_1>$.

The packing morphology is analyzed by calculating the radial distribution function (RDF) which is given by Eq. (9). The RDF can be interpreted as the number of particles that are located in a differential volume shell (thickness Δr) with a distance $r$ from the particle center, divided by the expected number of particles based on the particle number density of the whole packing. Here the center of mass of a cross- or star-shaped particle is considered as the particle center. The RDF plots show $G(r)$ over $r/L$, where $r$ has been normalized by the respective particle length $L$.

$$G(r) = \frac{N_{RDF}(r)}{4r^2 \Delta r N \rho_N} \qquad (9)$$

In Eq. (9) $\rho_N$ is the average number density of the particles (number of particles in the simulation domain divided its volume) and $N_{RDF}$ is the number of particles in the differential volume shell given by:

$$N_{RDF} = \sum_{i}^{N_{10L}} \sum_{j \neq i}^{N} \delta(r - r_{ij}) \tag{10}$$

where δ is the Dirac delta function, $r_{ij}$ is the distance between the center of particle i to the center of particle j and $N_{10L}$ is the number of particles which have a distance of at least 10 × L from each side wall as well as the top and bottom of the 2D domain. This area is sketched in Figure 4b. In the 3D case, $N_{10L}$ corresponds to the number of particles that have at least a distance of 10 × L from the top and bottom wall and a distance of 3 × L from the cylinder wall. We exclude particles close to the wall because these particles have no close neighbors outside of the walls.

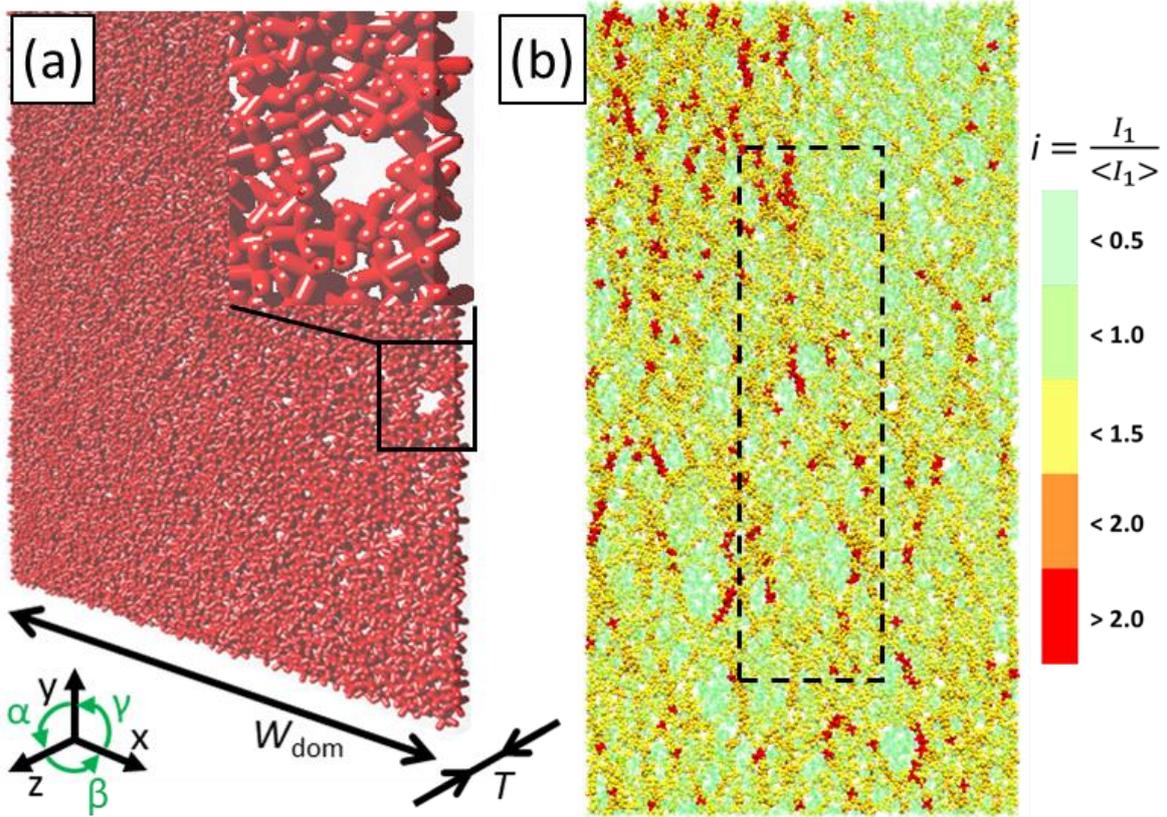

Figure 4: a) Visualization of a pseudo-2D packing of star-shaped particles (Ψ = 0.619) with $W_{dom}$ = 30 × L and T = L (L = particle length). b) Visualization of the normalized, first stress invariant (i = $I_{1,i}$/<$I_1$>) of the packing shown in (a). The dashed area depicts the area which is used to calculate the radial distribution function (RDF). The results of the RDF are shown in Figure 7.

## Results

### Contact force distributions

As a first approach, we assess whether our simulations support a previously reported trend that the exponential tail of P(f) becomes longer for increasingly non-spherical particles [13–15,26]. Currently, it is unclear whether such a trend is limited to a given domain dimensionality (2D vs 3D) and whether compressed packings show the same behavior as sheared systems. Figure 5 plots the probability distributions of the normal contact forces normalized by their mean, P(f), for different particle sphericities and domain configurations.

For 2D domains, the P(f) of cross-shaped particles with the highest investigated sphericity (Ψ = 0.995) and spheres are very similar (Figure 5a). In 2D packings, changes in the P(f) when transitioning from packings of spherical particles to slightly non-spherical particles seem to occur gradually. Figure 5a also plots Eq. (1), the numerical expression for *P(f)* proposed by Mueth et al. [12]. Somewhat surprisingly, the empirical Eq. (1), although extracted originally from a packing of spheres agrees well with the data for low sphericity cross-shaped particles in 2D, but not with the simulation results of spheres. A possible reason for this deviation might be the fact that Mueth et al. [12] studied 3D packings and only recorded contact forces at the wall. Indeed, the numerical results of a packing of spheres in a 3D domain, Figure 5d, agree very well with Eq. (1).

When comparing the contact force distribution, *P(f)*, of cross-shaped particles (2D domain) as a function of their sphericity (Figure 5a), one can observe that the length of the exponential tail of *P(f)* increases with decreasing $\Psi$. However, the increase in the length of the exponential tail is limited. Even in packings of cross-shaped particles with a sphericity of $\Psi$ = 0.473 (lowest sphericity modeled) only four individual contact forces (out of $10^5$) have a value of *f* > 8. The general type A shape of the distributions, i.e. a peak at *f* ≈ 1 and an exponential tail is independent of particle sphericity when the packing is restricted to 2D. Hence, our results confirm the observations by Saint-Cyr et al. [14] who modeled trimers (1 ≥ $\Psi$ ≥ 0.76) and observed that the shape of *P(f)* and the location of its peak is independent of particle sphericity in compressed 2D packings.

Turning now to pseudo-2D packings of star-shaped particles (Figure 5b and Figure 5c for, respectively, *T* = *L* and *T* = 2 × *L*), one observes a change in the shape of *P(f)* from type A to type B with decreasing particle sphericity. Generally, the peak at *f* ≈ 1 becomes less pronounced and the length of the exponential tail increases with decreasing sphericity. Only for particles with the highest sphericity ($\Psi$ = 0.995) the shape of *P(f)* in pseudo-2D packings coincides with the shape that is observed in 2D simulations.

When increasing the transverse thickness of the pseudo-2D simulations (*T* = 2 × *L*, Figure 5c) and ultimately reaching full 3D simulations, Figure 5d, the shape of *P(f)* changes further, i.e. the peak of *P(f)* remains at *f* ≈ 1 for high-sphericity particles but is no longer visible for low-sphericity particles ($\Psi$ = 0.461 and $\Psi$ = 0.419). Furthermore, the length of the exponential tail increases significantly for $\Psi$ < 0.7 when increasing the transverse thickess of the domain to *T* = 2 × *L* and 3D. It has been suggested that type A distributions are essentially Gaussian-like (centered around *f* ≈ 1, albeit truncated at *f* = 0) which would indicate that the forces are distributed homogeneously [13,18]. The similarity between a type A distribution and a Gaussian-shaped distribution is shown in Figure 1. One can see that a Gaussian is a good fit for type A distributions for *f* < 3, but the type A distribution has a longer tail. On the other hand, the lack of a peak at *f* ≈ 1 and the long exponential tails of type B distributions represent a heterogeneous force distribution with a large number of below-average contact forces but also some contact forces that are ten times above average.

To summarize, our results show that in pseudo-2D and 3D packings the shape of *P(f)* changes from type A to a type B when the particle sphericity decreases below the critical value $\Psi_{crit}$ = 0.7 (but not in 2D packings). The shape change comes with an increasing length of the exponential tail of *P(f)* and a decreasing prominence of the peak at *f* ≈ 1 with decreasing particle sphericity. Our results connect for the first time the observations of several previous studies: Saint-Cyr et al. [14] and Azéma and Radjai [15] who observed exclusively type A distributions in compressed 2D packings of non-spherical particles. On the other hand, Gan et al. [26] who simulated ellipsoids with $\Psi$ ≥ 0.7 in 3D packings observed type A distributions and Murphy et al. [13] who simulated Z-shaped particles with $\Psi$ = 0.45 in 3D packings observed type B distributions.

Combing the results of our simulations with previously reported observations allows us to draw the following general conclusion for the shape of *P(f)* in 2D and 3D packings of compressed non-spherical particles with different shapes and sphericities:

- With decreasing particle sphericity, the contact force distribution of compressed 3D packings becomes more heterogeneous. This is evidenced by the increasing length of the exponential tail of the contact force distribution with decreasing particle sphericity for $\Psi < \Psi_{crit} = 0.7$, independent of the specific particle shape.
- In compressed 2D packings the length of the exponential tail of the contact force distributions does not depend on particle sphericity.

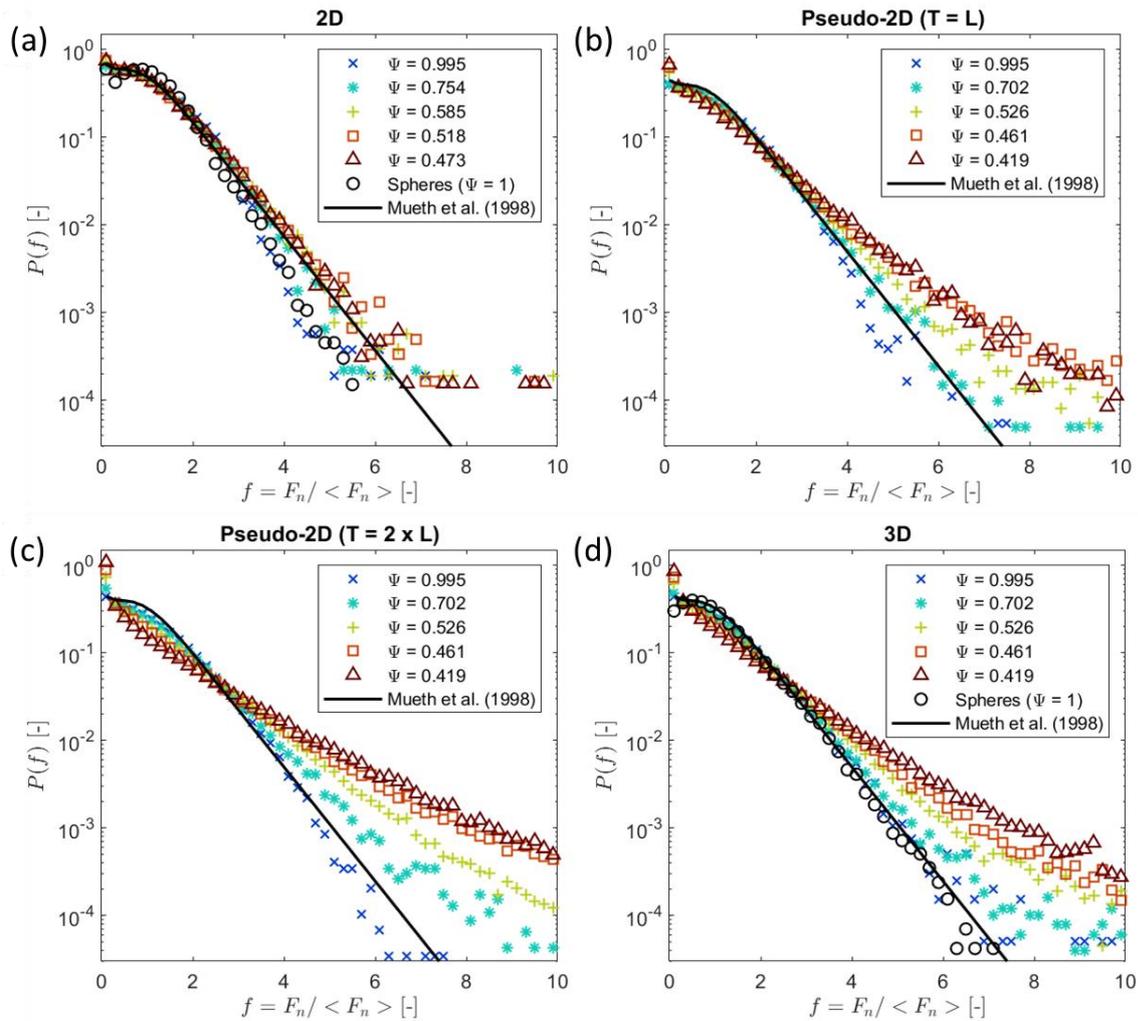

*Figure 5*: Probability distribution functions, $P(f)$, of the normal contact force ($F_n$) normalized by the mean normal contact force ($<(F_n)>$) for all four packing domains simulated. In the 2D domain (a) the flat, cross-shaped particles can only translate in the x- and y-directions. In the pseudo-2D domains (b) three-dimensional particles can translate in all three directions, with the transversal domain width $T$ in the z-direction being as large as the particles ($T = L$). In the second pseudo-2D domain (c) the z-direction is twice as large as the particles ($T = 2 \times L$). The 3D domain (d) is cylindrical with diameter $D_d = 10 \times L$. For each packing configuration the full range of particle sphericities simulated is shown. For reference a 2D simulation of spheres is included in (a) and a 3D simulation of spheres is included in (d). Each panel also plots the probability distribution function predicted by the empirical equation (Eq. (1)) of Mueth et al. [12], which was derived from compressed 3D packings of glass spheres.

## Quantification of the length of the exponential tail

Next, a quantitative description of the length of the exponential tail is explored. In the following, we will focus on the region $f > 1$ of $P(f)$, due to the importance of the large forces which can potentially lead to the fracture of particles and endanger a stable packing.

Figure 6a plots the exponent $k$ as a function of particle sphericity. The exponent $k$ is obtained by fitting the exponential tail ($f > 1$) of the different $P(f)$ (Figure 5) with Eq. (2). Error bars are omitted in Figure 6a for clarity (the 95% confidence bounds for $k$ are typically within ±0.025). Qualitatively the homogeneity of $P(f)$ decreases as the length of the tail increases (i.e. a lower value of $k$).

For particles of high sphericity ($\Psi = 0.995$), the values obtained for $k$ are in the range $1.28 \geq k \geq 1.47$. These values are in between the values obtained by Mueth et al. [12] (packings of glass spheres in 3D with $k = 1.5$) and Gan et al. [26] (3D packing of spheres with $k = 1.24$).

Turning now to less spherical particles: In 2D packings, $k$ decreases slightly with decreasing $\Psi$, i.e. $k = 1.42$ for $\Psi = 0.995$ and $k = 1.32$ for $\Psi = 0.47$. In 2D packing the decrease of k with decreasing $\Psi$ can be fitted well by a linear function (dashed black line in Figure 6a). The comparatively high value of $k = 1.32$ for the lowest sphericity values studied ($\Psi = 0.47$) shows that there are relatively few cases of high contact forces in 2D packing of low sphericity particles. Good agreement is also seen when including data for $k$ obtained by other works that have assessed compressed 2D packings (e.g. the data of Saint-Cyr et al. [14] is for trimer particles).

In pseudo-2D and 3D packings $k = 1.38 \pm 0.14$ for $\Psi > \Psi_{crit} = 0.7$. For $\Psi < \Psi_{crit}$ $k$ decreases exponentially with decreasing $\Psi$ and reaches a value of $k = 0.87 \pm 0.13$ for the lowest sphericity investigated, i.e. $\Psi = 0.42$. An exponential fit of $k$ in pseudo-2D ($T = 2 \times L$) packings is shown by a solid line in Figure 6a. Hence, for $\Psi < \Psi_{crit}$ the probability of finding large contact forces ($f > 8$) increases exponentially and the contact force distributions become increasingly heterogeneous. The critical sphericity value, i.e. $\Psi_{crit} = 0.75$ for cross-shaped particles (2D) and $\Psi_{crit} = 0.7$ for star-shaped particles (3D) is the lowest sphericity for which a contact between two particles always involves the hemispherically-capped ends of the particles. This can be explained by the fact that, at this critical sphericity the arms protruding from a particle are exactly twice as long as the particle radius (Figure 6c). For $\Psi > \Psi_{crit}$, contacts will always involve the end-caps of a particle (Figure 6b), while for $\Psi < \Psi_{crit}$, contacts can also involve the flat/cylindrical section of the protruding arms (Figure 6c). Particles that only contact each other with the hemispherical end-caps, i.e. star-shaped particles with $\Psi \geq 0.7$, are more likely to slip relative to each other when a load is applied. Conversely, particles with contacts that involve the flat/cylindrical sections of the arms, i.e. star-shaped particles with $\Psi < 0.7$, are less likely to slip relative to each other which means that they are more likely to jam. When a particle jams during compression, the contact forces acting on such a particle can increase substantially (and without the particle unjamming the contact forces cannot relax). These high contact forces give rise to the long exponential tail of $P(f)$ for low sphericity particles, in particular for star-shaped particles with $\Psi < 0.7$. Whereas this rationale explains the transition at $\Psi_{crit} = 0.7$ for pseudo-2D and 3D packings, it is unclear why such a pronounced transition is absent for 2D packings. We speculate that the reason might lie in the particular spatial distribution (morphology) of the particles which will be investigated in the following.

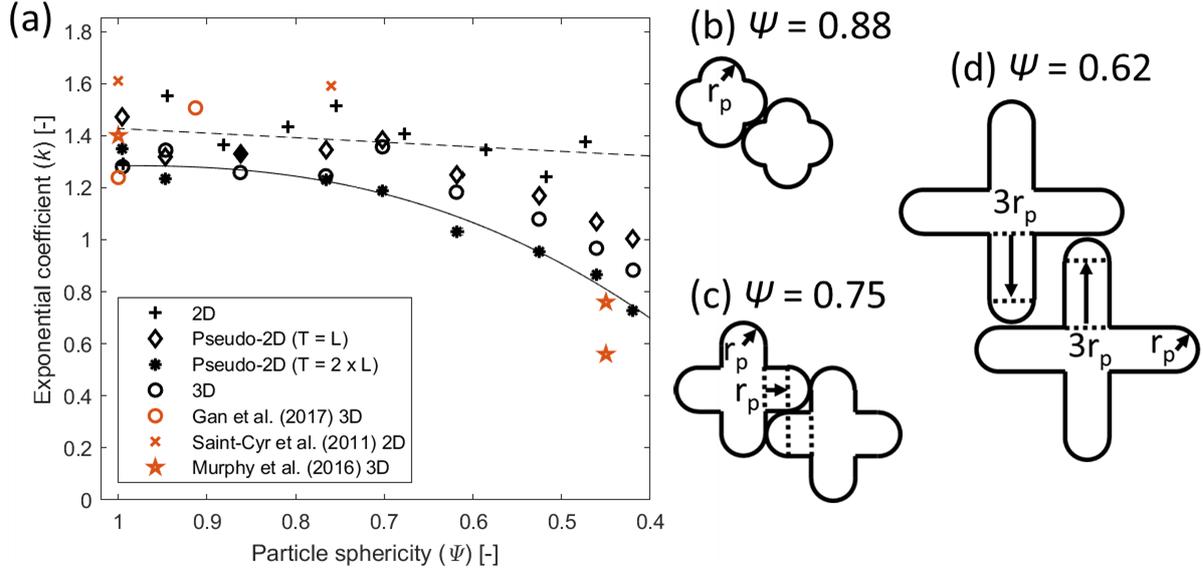

Figure 6: (a) Exponent $k$ (Eq. (2)) obtained by fitting the exponential tail ($f > 1$) of the contact force distributions. Results of the present work are shown by black markers, while red markers denote results from previously published studies. The dashed line is a linear fit ($k(\Psi) = 0.18(1-\Psi)+1.43$) to the values of $k$ obtained in 2D packings (+), while the solid line is an exponential fit ($k(\Psi) = -2.03(1-\Psi)^{2.43}+1.29$) to the values of $k$ obtained in pseudo-2D ($T = 2 \times L$) packings (*). b)-d) Sketches of cross-shaped particles showing how the length of the protruding arms change with sphericity.

## Packing morphology

By analyzing the probability distribution of the contact forces, $P(f)$, we found that in compressed pseudo-2D and 3D particle packings the length of the exponential tail of $P(f)$ increases with decreasing particle sphericity. In contrast, in 2D packings, the length of the exponential tail of $P(f)$ does not depend strongly on the particle sphericity. Hence, the question why the behavior of 3D packings differs distinctively from 2D packings remains unanswered. Instead of assessing the structure of a packing by an averaged, global parameter, such as the solid fraction, the morphology of the packings is assessed by calculating their radial distribution functions, RDF ($G(r)$, Figure 7).

Comparing the RDFs in 2D packings as a function of sphericity (Figure 7a), one observes that for cross-shaped particles of low sphericity ($\Psi = 0.75$) the first ($r/L = 0.76$) and second ($r/L = 1.5$) peak, are more prominent compared to the first ($r/L = 1$) and second ($r/L = 1.91$) peak of more spherical particles ($\Psi = 0.995$). These peaks correspond to particle configurations of local crystallinity which are also the closest possible arrangements of interlocking particles. Sketches of such crystalline particle configurations are shown in Figure 7 (and enhanced in the Supplemental Material [43]). More pronounced peaks imply a more frequent occurrence of the respective particle configurations. The distances $D_1$, $D_2$, $D_3$ between the centers of two interlocking particles, as defined in Figure 7, can be calculated by:

$$D_1 = \sqrt{d_p^2 + [(L + d_p)/2]^2}, \qquad D_2 = 2D_1, \qquad D_3 = 3D_1 \qquad (11)$$

For cross-shaped particles with $\Psi = 0.75$ the values of the first ($r/L = 0.76$) and second ($r/L = 1.5$) peak of the RDF are close to the geometrically determined values of the $D_1$ and $D_2$ configurations ($D_1/L = 0.75$ and $D_2/L = 1.49$). This indicates that the peaks in the RDF indeed correspond to the proposed closest crystalline configurations. In the RDF of 2D packings of particles with $\Psi = 0.75$ the peak positions are shifted to lower values of $r/L$ compared to the RDF of 2D packings of more spherical particles ($\Psi = 0.995$). The position of the peaks ($D_1/L$) shifts to lower values for decreasing $\Psi$ because $L$ increases with decreasing $\Psi$ and $\frac{D_1}{L} \propto \frac{(L+\sqrt{L})}{2L} < 1$. For 2D packings of cross-shaped particles with $\Psi = 0.75$ a small third peak can be seen at $r/L = 2.23$, which corresponds to a similar packing configuration as the one described above (four-particle configuration) with an analytical value of $D_3/L = 2.24$. These first three peaks can also be seen in the RDFs of 2D packings of particles with lower sphericity. The positions of the peaks as well as the analytically obtained positions are shown in Table 2. Since these three peaks are observed even in 2D packings of highly non-spherical particles one can conclude that structures of local crystallinity can be observed in all of these packings.

Table 2: Peak positions observed in the RDFs of 2D packings of cross-shaped particles and peak positions calculated according to Eq. (11).

| $\Psi$ | 1ST PEAK | $D_1/L$ | 2ND PEAK | $D_2/L$ | 3RD PEAK | $D_3/L$ |
|---|---|---|---|---|---|---|
| **0.473** | 0.55 | 0.55 | 1.12 | 1.11 | 1.66 | 1.66 |
| **0.518** | 0.58 | 0.57 | 1.14 | 1.14 | 1.73 | 1.71 |
| **0.585** | 0.62 | 0.61 | 1.25 | 1.21 | 1.82 | 1.82 |
| **0.754** | 0.76 | 0.75 | 1.50 | 1.49 | 2.23 | 2.24 |
| **0.995** | 1.00 | 0.98 | 1.91 | 1.95 | none | 2.93 |

Figure 7b plots the RDFs of star-shaped particles in 3D packings. The RDF of particles with $\Psi = 0.995$ shows a first peak at $r/L = 1$ and a second peak at $r/L = 1.9$. These peaks are at the same positions as in the RDF of 2D packings of cross-shaped particles with $\Psi = 0.995$ because both particle shapes are almost spherical and cannot interlock. In 3D packings of star-shaped particles with $\Psi = 0.702$ the first peak is located at $r/L = 0.7$ and the second peak at $r/L = 1.37$. Compared to 2D packings of cross-shaped particles with similar sphericity, the peaks are shifted to lower $r/L$ values. This observation can be explained by the fact that star-shaped particles in 3D packings have additional degrees of freedom compared to cross-shaped particles in 2D packings allowing closer packing configurations in 3D packings. A sketch of the closest packing of star-shaped particles in 3D is shown in Figure 7b. The closest distance ($J_1$) between the centers of two star-shaped particles in the packing configuration shown in Figure 7b is:

$$J_1 = \sqrt{3} d_p \qquad (12)$$

Hence, in 3D packings of star-shaped particles with $\Psi = 0.702$, the first peak $J_1$ is expected at $r/L = 0.58$ and the second peak at $r/L = 1.15$. However, the first and second peak are found at $r/L = 0.7$ and $r/L = 1.37$, respectively. This result indicates that the most likely packing configuration of star-shaped particles in a compressed 3D packing is considerably looser than the closest possible crystalline packing configuration. Hence, 3D packings have a different morphology to 2D packings, as the peak location in the RDF of 2D packings of cross-shaped particles agrees very well with the closest possible crystalline packing configuration which, however, is not the case for 3D packings.

Additionally, when comparing the RDF of 2D and 3D packings, one finds that the height of the peaks in the RDF of 3D packings is lower than in 2D packings, a further sign of a reduced crystallinity when introducing an additional dimension. The reduced peak height is particularly noticeable when comparing the RDF of cross-shaped particles with $\Psi = 0.754$ (2D packing, Figure 7a) with the RDF of star-shaped particles with $\Psi = 0.702$ (3D packings, Figure 7b). Even a third peak is visible in the RDF of the 2D packing ($\Psi = 0.754$), whereas a third peak is absent in the RDF of a 3D packing ($\Psi = 0.754$). The reduction in crystallinity in 3D packings is more pronounced for particles of lower sphericity, i.e. $\Psi <$ 0.702 ($\Psi < \Psi_{crit}$); in such packings even the first peak in the RDF disappears completely, indicating an amorphous packing structure. A reduced peak height in the RDF of packings of rod-like particles with decreasing particle sphericity has also been observed in previous works [53–55].

To summarize, the crystallinity of 3D packings of star-shaped particles decreases with decreasing particle sphericity. In 2D packings of cross-shaped particles, however, such a decrease in crystallinity with decreasing sphericity could not be observed. The data in Figure 7 shows that interlocked, crystalline, configurations are found in 2D packings of cross-shaped particles with $\Psi$ < 0.7, whereas such configurations do not seem to be present to a large extend in 3D packings of star-shaped particles with $\Psi$ < 0.7. This can be explained by considering that 3D star-shaped particles have three additional degrees of freedom (one translational degree and two rotational), compared to 2D cross-shaped particles. It is therefore less likely that star-shaped particles, when dropped into a 3D container followed by compression will configure themselves into a highly crystalline packing.

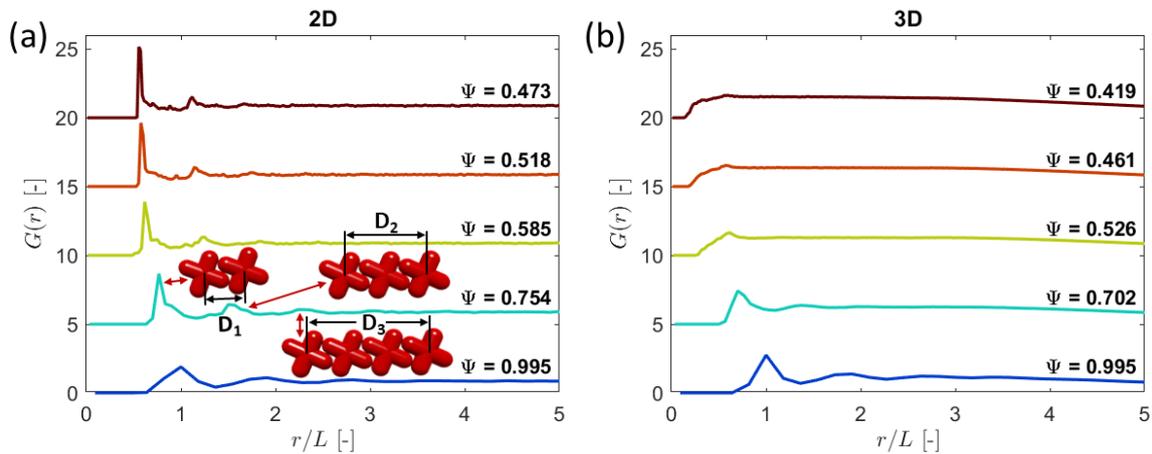

Figure 7: Radial distribution function (RDF), G(r), for (a) 2D and (b) 3D packings of particles of different sphericity. For better readability, the curves are shifted vertically with an offset of 5. In 2D packings, the distance between the particle centers corresponding to the first, second and third peak are labeled $D_1$, $D_2$ and $D_3$. The distances $D_{1-3}$ can be determined analytically according to Eq. (11), and $J_1$ according to Eq. (12).

## Linking force distributions to packing morphology

We described above that in 3D packings, below a critical value of $\Psi < \Psi_{crit} = 0.7$, the length of the exponential tail of the contact force distribution increases with decreasing particle sphericity. The increasing length of the exponential tail in 3D packings (and not in 2D packings) for sphericities $\Psi$ < 0.702 seems to coincide with the disappearance of peaks in the RDF. To understand this observation, we revisit the work by Luding [17], who used the DEM to investigate particle stresses of monodisperse and polydisperse packings of discs in a perfectly hexagonal 2D lattice. For perfectly monodisperse particles, a uniform particle stress distribution was observed. However, as soon as some polydispersity was introduced, by varying the disc diameter by a small amount (±0.33%), localized force chains were observed in the particle packing. Luding [17] also observed that the probability of large particle stresses to occur increases with increasing polydispersity, i.e. the length of the exponential tail of the particle stress distribution increases with increasing polydispersity. One can interpret the findings of the present work as complement to Luding's [17] results for non-spherical particles, i.e. a decreasing crystallinity of the packing (induced by either polydispersity or non-sphericity) leads to wider, less homogeneous, contact force and particle stress distributions, provided that the following two assumptions hold: (1) The introduction of polydispersity does lead to a reduction in crystallinity and (2) the behavior of the particle stress distribution, $P(i)$, depending on particle sphericity is very similar to that of the contact force distributions, $P(f)$.

Concerning the first assumption, there is indeed evidence of reduced peak heights in the RDF of a hexagonal packing with 5% polydispersity, when compared to the RDF of a monodisperse packing [56].

The reduced peak height hints towards a reduced crystallinity in polydisperse packings, however, further research is required to confirm the assumption.

To confirm the second assumption, that the exponential decay of the particle stress distribution is similar to the decay of the contact force distribution, one can compare $P(f)$ and $P(i)$, i.e. the probability distribution of the normalized first stress invariant $i = I_{1,i}/\langle I_1 \rangle$ shown in Figure 8. Generally, the shape of $P(i)$ is similar to the shape of $P(f)$, i.e. for 2D packings distributions of type A and for 3D packings of low sphericity particles distributions of type B are obtained. For high sphericity ($\Psi > 0.7$) cross-shaped particles in 2D packings (Figure 8a), the distributions exhibit a pronounced peak at $i = 1$, which disappears for $\Psi < \Psi_{crit} = 0.7$ as $\lim_{i \to 0} P(i)$ increases with decreasing $\Psi$. Specifically, in 3D packings of star-shaped particles, $P(i)$ has a type A-like distribution for $\Psi \geq 0.7$ (peak at $i = 1$ and an exponential tail of similar length as in 2D packings), and no particle experiences a stress invariant with a magnitude of more than 6 times the mean. However, for 3D-packings of star-shaped particles with $\Psi < \Psi_{crit}$, the shape of $P(i)$ changes to a type B distribution and the length of the exponential tail of $P(i)$ increases with decreasing particle sphericity. The steep increase of $\lim_{i \to 0} P(i)$ implies that there is an increasing number of particles that experience only a very small fraction of the load that is put on the packing by uniaxial compression. At the same time, owing to the increasing length of the exponential tail, some particles experience stresses that are significantly higher than the mean. These trends match the behavior of $P(f)$, as described further above.

The transition of $P(i)$ to a type B distribution in 3D packings of star-shaped particles for decreasing sphericity, is attributed to an increasing frequency of contacts between the flat parts of the arms protruding from the particles (Figure 6c). Such contacts are only possible for star-shaped particles with $\Psi < \Psi_{crit}$, whereas for $\Psi \geq \Psi_{crit}$ all contacts between star-shaped particles involve the hemispherical end-caps of the spherocylinders. Particles forming contacts between the flat parts of the spherocylinders are less likely to rearrange during uniaxial compression. This effect can also be interpreted as an increased apparent friction coefficient between particles. Such particles are more likely to jam during uniaxial compression, instead of rearranging into a configuration which would reduce the stress acting on the particle. As a consequence, high stresses can build up which results in an increased length of the tail of $P(i)$. The build-up of stresses in some particles leaves other particles to contribute little to the stress transmission in the packing leading in turn to an increase of $\lim_{i \to 0} P(i)$ with increasing apparent friction [57].

To summarize, the dependence of the length of the exponential tail of $P(i)$ on particle sphericity is very similar to the respective behavior of $P(f)$. Hence, our results can indeed be considered as complementary to Luding's [17] observation on the effect of polydispersity on the stress distribution in particle packings, i.e. a decreasing crystallinity (due to increasing non-sphericity or polydispersity) leads to a extended tail of the contact force and particle stress distributions.

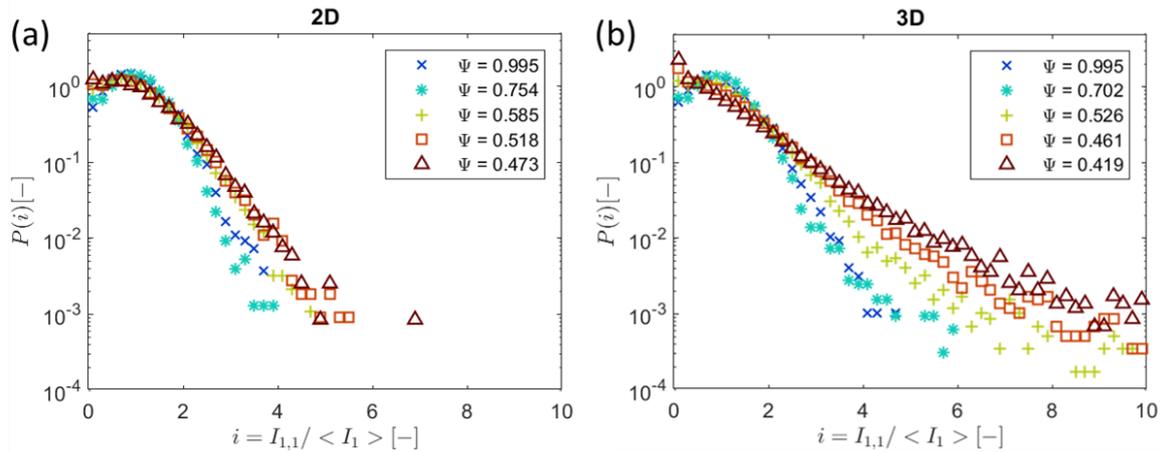

*Figure 8: Probability distributions of the normalized first stress invariants P(i) as a function of particle sphericity in (a) 2D and (b) 3D packings. The normalized first stress invariants i are obtained by normalizing the first stress invariants $I_{1,i}$ by $<I_1>$.*

## Conclusions

This work investigated compressed 2D and 3D packings of non-spherical, non-convex, cross- and star-shaped particles using a newly developed DEM. Such non-convex particles can interlock forming packings that may sustain compressive and tensile forces despite containing purely non-cohesive particles. The particle sphericity ($\Psi$) was varied in the range $\Psi$ = 0.42 - 1. The morphology of the packings was investigated by calculating the radial distribution function (RDF). Through the RDF we have established a link between the packing morphologys and the contact force probability distributions as expressed by the decay exponent *k*.

In 2D packings of cross-shaped particles a linear decrease of *k* was observed, from *k* = 1.42 for $\Psi$ = 0.995 to *k* = 1.32 for $\Psi$ = 0.47. For pseudo-2D and 3D packings of star-shaped particles *k* is independent of the sphericity (*k* = 1.38±0.14) for $\Psi \geq 0.7$; however, for smaller sphericities (i.e. $\Psi$ < 0.7) the magnitude of *k* decreases exponentially with decreasing $\Psi$. These findings connect for the first time the results of previous works [13,14,26] on compressed packings of non-spherical particles. This allows us to establish the following general correlations between $\Psi$ and the heterogeneity of the distribution of the contact forces expressed by the magnitude of *k*:

- In compressed 3D packings, the distribution of contact forces in the normal direction becomes increasingly heterogeneous, i.e. the length of the exponential tail increases exponentially with decreasing particle sphericity, independent of the specific particle shape.
- In 2D packings the influence of particle sphericity on the distribution of the contact force distribution in the normal direction is small. In contrast to the exponentially decreasing *k* with decreasing $\Psi$ in 3D packings, the decrease of *k* with decreasing $\Psi$ in 2D packings is linear.

The heterogeneity of the contact force distribution needs to be considered when designing particles for specific applications such as aleatory construction for which non-convex particles with low sphericity (e.g. $\Psi$ = 0.45 [13]) are desired. Such particles need to be able to withstand the highest contact forces, that can reach values that are an order of magnitude higher than the mean contact force.

We further establish a link between the increasing heterogeneity of the distribution of the contact forces and the packing morphology in packings of non-spherical particles. We have demonstrated that the increasing heterogeneity in the contact force distributions with decreasing sphericity correlates with a decreasing crystallinity of the packings. The link between a decreasing packing crystallinity and more heterogeneous contact force distributions has been postulated previously by Luding [17] for spherical particles, when assessing the effect of polydispersity on the homogeneity of the particle stress distributions. Hence, our results can be interpreted as complementary to this previous observation providing further evidence that a reduced packing crystallinity, through either an increase of domain dimensionality, particle non-sphericity or polydispersity, leads to a more heterogeneous stress distribution.

## Acknowledgments

This work has been financed by the Swiss National Science Foundation grant No. 200021_157122/1 and 200020_182692.

# Supplemental Material to "The Link Between Packing Morphology and the Distribution of Contact Forces and Stresses in Packings of Highly Non-Convex Particles"

N. A. Conzelmann[1,2], A. Penn[1], M. N. Partl[2], F. J. Clemens[2], L. D. Poulikakos[2], C. R. Müller[1]

[1] ETH Zürich, Laboratory of Energy Science and Engineering, Department of Mechanical and Process Engineering, Institute of Energy and Process Engineering, Leonhardstrasse 21, 8092 Zürich, Switzerland

[2] Empa – Swiss Federal Laboratories for Materials Science and Technology, Überlandstrasse 129, 8600 Dübendorf, Switzerland

## I. Packing fraction

Particles of low sphericity form packings with a low solid fraction. Figure S1 plots the solid fraction of a packing as a function of $\Psi$ and domain geometry. For 2D packings, we show the area fraction occupied by particles as well as the volume fraction, assuming that the domain has a transverse thickness that is equal to the particle diameter ($T = d_p$).

For a packing of non-spherical particles ($\Psi > 0.9$), one observes an increase in solid fraction with decreasing sphericity, with a peak at $\Psi = 0.9 - 0.95$. At this sphericity value, the particle shapes are the closest to a cuboidal shape. Cuboids can be stacked without any gaps. Reducing the sphericity further ($\Psi \leq 0.9$), the particles become increasingly concave with an increased tendency to interlock, which leads to a decreasing solid fraction of the packing. The shape of the solid fraction versus sphericity curves is similar to the trends that have been observed previously simple spherocylinders and ellipsoids [1–3].

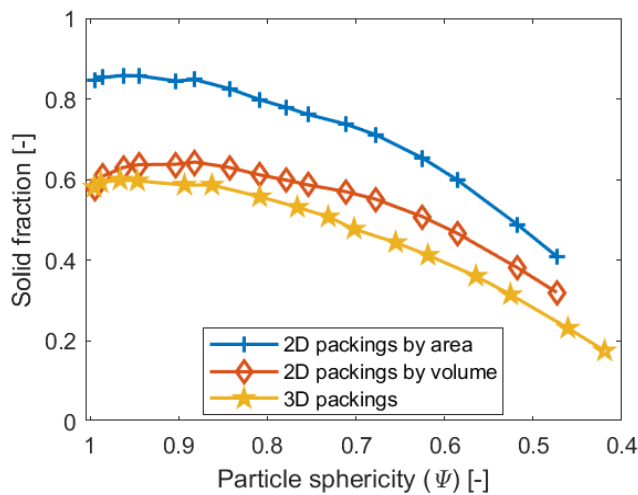

Figure S1: The solid fraction of packings as a function of sphericity and domain geometry.

## II. Closest possible particle configurations

The peaks in the RDF, shown in **Error! Reference source not found.** in the main text, result from specific particle configurations in which particles have the shortest possible distance between adjacent particle centers. Figure S2 visualizes such particle packing configurations for cross-shaped and star-shaped particles.

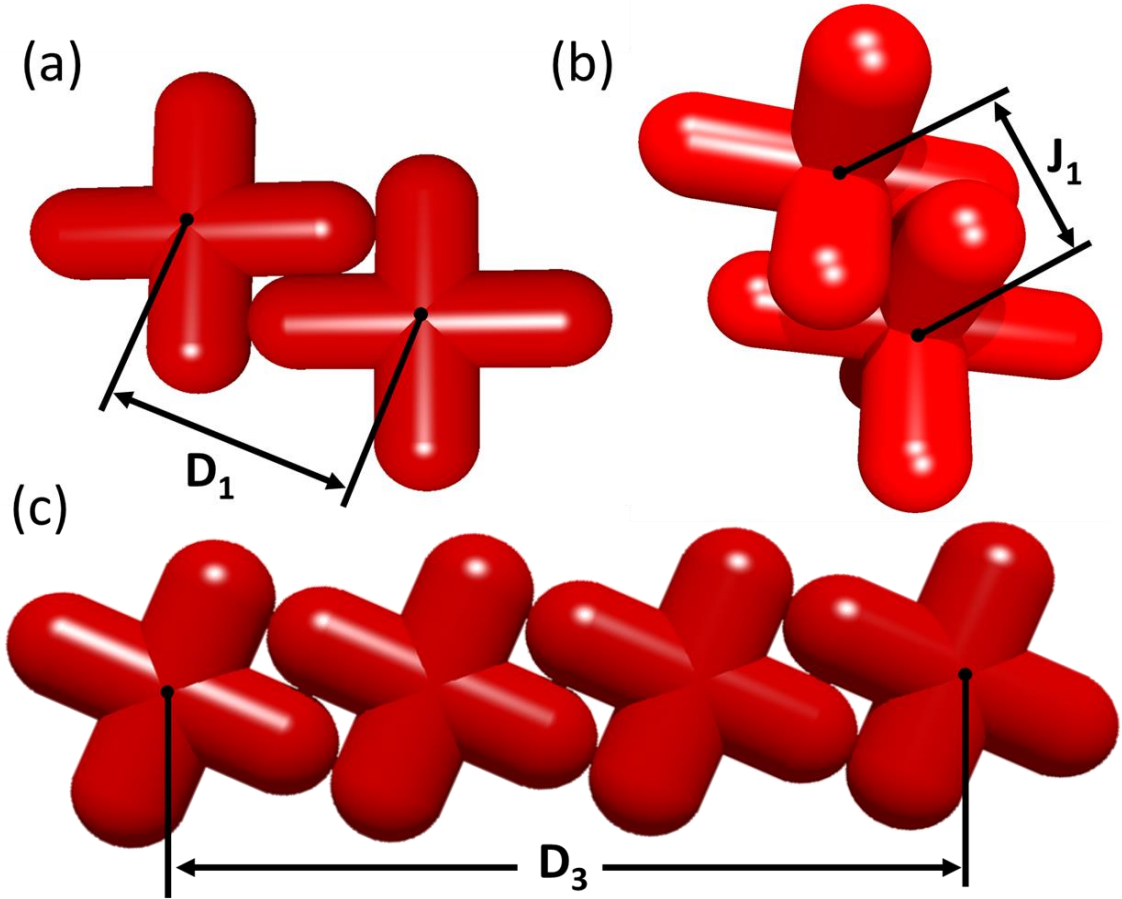

Figure S2: Visualization of particle configurations with the shortest possible distance ($D_1$, $D_3$, $J_1$) between particle centers (black dots): (a) cross-shaped particles used in 2D simulations, (b) star-shaped particles used in pseudo-2D and 3D simulations and (c) configuration of four cross-shaped particles. $D_1$ and $D_3$ can be calculated according to equation Eq. (11), and $J_1$ according to Eq. (12).

## III. Conversion between particle sphericity and aspect ratio

In this work, the particle shapes are identified by the particle sphericity, while some other works identify particles of the same star-shape by the aspect ratio, which is defined as the overall length of a particle ($L$) and the diameter of the protruding arms ($d_p$) [4,5]. To aid the comparison to other works, Table S1 shows the relationship between the particle aspect ratio and particle sphericity.

Table S1: Relationship between the particle aspect ratio and the particle sphericity. The particle aspect ratio is defined as the overall length of a particle (L) and the diameter of the protruding arms ($d_p$).

| Particle aspect ratio ($L/d_p$) | Sphericity cross-shape (2D) | Sphericity star-shape (3D) |
|---|---|---|
| 1.1 | 0.995 | 0.995 |
| 1.7 | 0.933 | 0.933 |
| 2.25 | 0.842 | 0.809 |
| 3 | 0.754 | 0.702 |
| 4 | 0.677 | 0.619 |
| 6 | 0.585 | 0.526 |
| 8.5 | 0.518 | 0.462 |
| 11 | 0.473 | 0.419 |

## IV. Influence of particle friction

The following section discusses the influence of the interparticle friction coefficient $\mu$ on the probability distribution of the contact force $P(f)$. Throughout the main body of this work $\mu = 0.5$ was used.

Figure S3a plots $P(f)$ for cross-shaped particles with the highest ($\Psi = 0.995$) and lowest ($\Psi = 0.518$) sphericity in a 2D domain for varying $\mu$. For both $\Psi = 0.995$ and $\Psi = 0.518$, $P(f)$ is not affected appreciably by the friction coefficient.

Figure S3**Error! Reference source not found.**b plots $P(f)$ for star-shaped particles in a 3D domain (cylinder with a diameter $D_d = 6 \times L$). For the highest sphericity ($\Psi = 0.995$), $P(f)$ is independent of $\mu$. For the lowest sphericity ($\Psi = 0.461$), the $P(f)$ for $\mu = 0.5$ and $\mu = 0.9$ coincide ($\mu = 0.5$ was used in the main body of this work). However, for particles with $\Psi = 0.461$ and $\mu = 0.1$, $P(f)$ has a shorter tail compared to $P(f)$ for higher values of $\mu$, indicating that there is a minor influence of $\mu$ on $P(f)$. A reason for this observation may be that it is less likely for low friction particles to obtain configurations in rotationally jammed configurations. Instead such low-friction particles are more likely to arrange themselves into the closest crystalline configuration which would be represented by a shorter exponential tail.

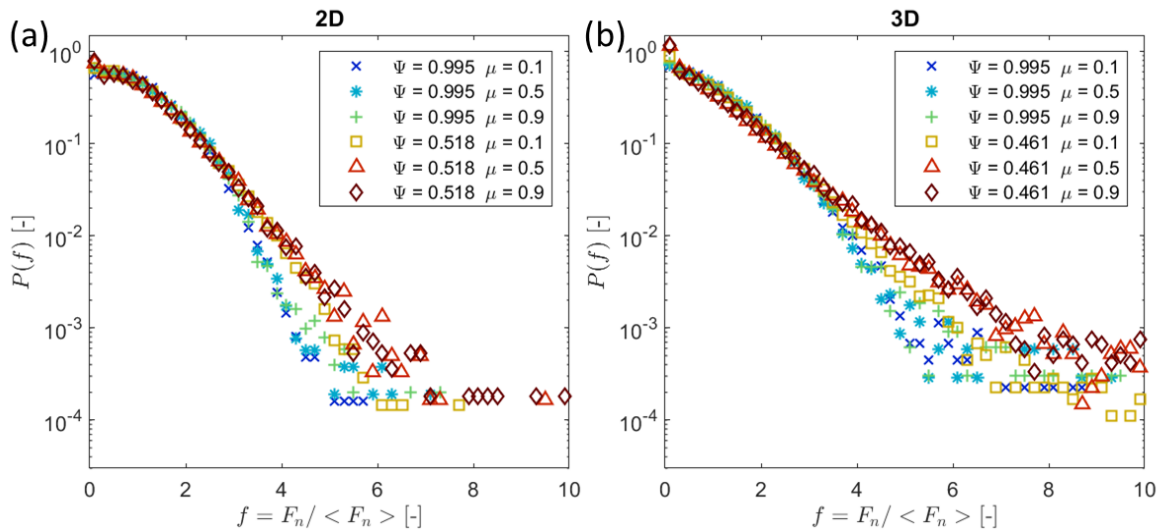

Figure S3: Probability distribution functions, $P(f)$, of the normal contact force ($F_n$) normalized by the mean normal contact force ($<(F_n)>$) for varying inter-particle friction coefficients $\mu$. The 2D data (a) uses flat cross-shaped particles, the pseudo-2D and 3D data (b) uses star-shaped particles.

Figure S4 probes further the influence of a low friction coefficient ($\mu = 0.1$) on $P(f)$. For star-shaped particles with $\Psi = 0.461$ in a 3D domain, the exponential tail of $P(f)$ is longer than for particles with $\Psi = 0.995$ in a 3D domain or any $P(f)$ in a 2D domain. Hence, the general trends observed in the main body of this work are not affected by a varying $\mu$, i.e. in 3D packings the length of the exponential tail increases with decreasing particle sphericity independent of $\mu$.

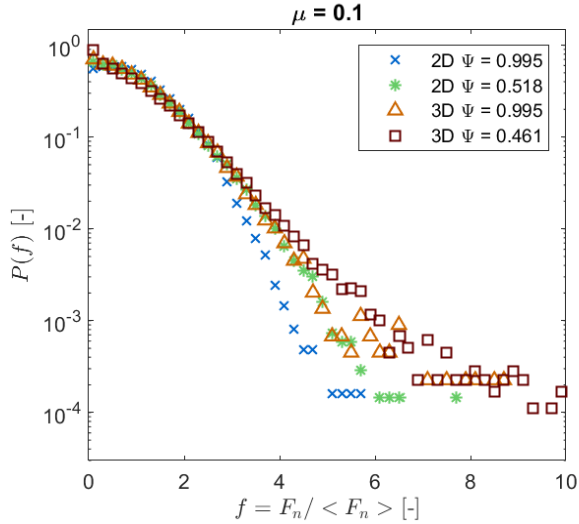

Figure S4: Probability distribution functions, *P(f)*, of the normal contact force ($F_n$) normalized by the mean normal contact force ($\langle F_n \rangle$) for an inter-particle friction coefficient $\mu = 0.1$. The particle sphericities plotted are the highest and lowest sphericities that have been considered in this work.

## V. Influence of wall friction

Friction between the particles and the domain walls is disregarded in this work to aid comparisons with previously reported numerical works of non-spherical particles [6,7]. In addition, other authors [8] have argued that introducing wall friction leads to more heterogeneous particle packings. In the following the influence of the wall friction coefficient $\mu_w$ on the probability distribution function of the normal contact force *P(f)* is assessed. An understanding of how $\mu_w$ influences *P(f)* will aid the comparison between experimental works and our numerical study.

Figure S5 plots *P(f)* for varying $\mu_w$ for particles with the highest sphericity parameter simulated in this work ($\Psi = 0.995$). For 2D simulations, $W_{dom}$ (x-direction) is equal to 30 times the particle length *L*; while pseudo-2D and 3D simulations are executed in smaller domains than the main body of the work to reduce computational time. Figure S5b plots the results obtained in a cuboidal domain with a transverse thickness *T* (z-direction) equal to *L* and $W_{dom} = 16 \times L$ (in the main body of this work $W_{dom} = 30 \times L$). Figure S5c plots the results obtained in pseudo-2D domains with $T = 2 \times L$ and $W_{dom} = 16 \times L$ ($W_{dom} = 30 \times L$ in the main body of this work). Figure S5d plots the results obtained in a 3D cylindrical domain with $D_d = 6 \times L$ ($D_d = 10 \times L$ in the main body of this work).

The results given in Figure S5a show that in a 2D domain *P(f)* does not vary appreciably with $\mu_w$. This is not unexpected as in a 2D domain particles have only contact with the side walls in the x-direction (comparatively large dimension of the domain in the x-direction with $W_{dom} = 30 \times L$). Hence, the domain width is sufficiently large to ensure frictional walls do not influence the packing. For pseudo-2D domains the shape of *P(f)* changes; i.e. a longer exponential tail is observed when wall friction is introduced. In a pseudo-2D packing, particles have (frictional) contacts with the side walls in the x-direction ($W_{dom} = 16 \times L$) and in the z-direction; however, the domain size in the z-direction is very small with $T = L$ (Figure S5b) and $T = 2 \times L$ (Figure S5c). Therefore, in pseudo-2D domains with $T = L$ every particle has a contact with at least one frictional wall and for $T = 2 \times L$ the majority of particles has at least one contact with a frictional wall. Due to the additional tangential contact forces between particles and frictional walls, it is more likely that particles become jammed during compression. These jammed particles cannot rearrange to distribute the contact forces more homogeneously. Interestingly, the exponential tails seem to be longer for $T = L$ than for $T = 2 \times L$. This can be explained by the fact that particles can barely move in the z-direction for $T = L$, while in a $T = 2 \times L$ domain two particles fit side-by-side in the z-direction and depending on their orientation a small gap can exist between the two particles into which a particle can wedge from above. Such a configuration would increase the normal forces between particles and the walls and consequently also the (limiting) tangential force that is given by Coulombs law. In a 3D domain (Figure S5d) the influence of frictional walls on P(f) is noticeable that is a longer exponential tail when introducing frictional walls. However, the influence of frictional walls in 3D domains is less pronounced than in pseudo-2D domains, since the diameter of the cylindrical domain is $D_d = 6 \times L$ and therefore less particles are in contact with the frictional wall when compared to the pseudo-2D domains. One can expect that if the diameter of the 3D domain is increased further the influence of the frictional walls on P(f) will decrease further. Additionally, the magnitude of $\mu_w$ does not seem to affect P(f) as long as $\mu_w > 0$.

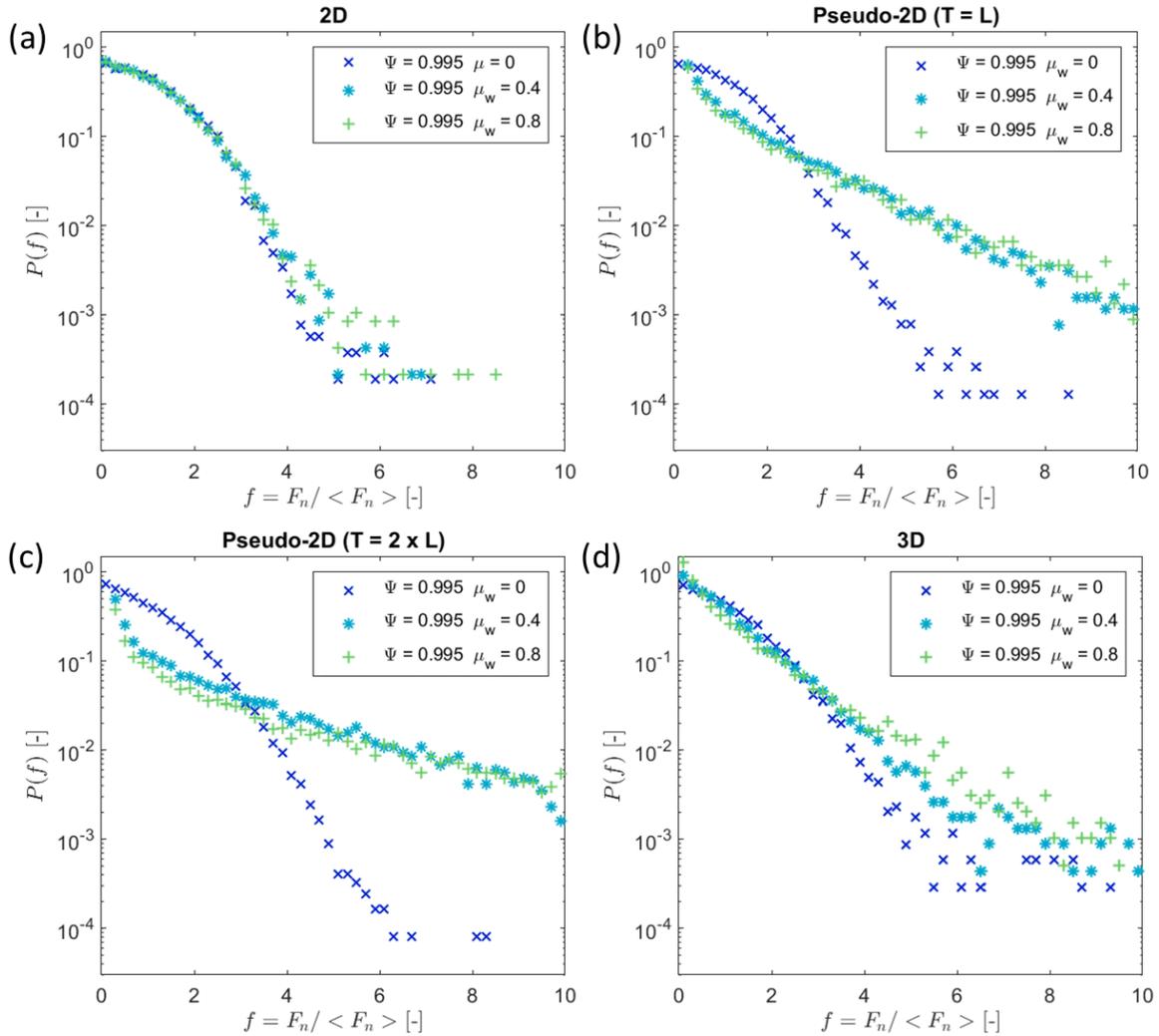

Figure S5: Probability distribution functions of the normal contact force, P(f) normalized by the mean normal contact force (<($F_n$)>) as a function of the wall friction coefficient $\mu_w$: (a) 2D domain, (b) pseudo-2D domain ($T = L$), (c) pseudo 2D domain ($T = 2 \times L$) and (d) 3D domain.